\documentclass[
]{ceurart}

\usepackage{listings}
\usepackage{subcaption}
\newcolumntype{P}[1]{>{\centering\arraybackslash}p{#1}}

\begin{document}

\copyrightyear{2025}
\copyrightclause{Copyright © 2025 for this paper by its authors. Use permitted under Creative Commons License Attribution 4.0 International (CC BY 4.0).}

\conference{LLM-TEXT2KG 2025: 4th International Workshop on LLM-Integrated Knowledge Graph Generation from Text (Text2KG), June 1 - June 5, 2025, co-located with the Extended Semantic Web Conference (ESWC 2025), Portorož, Slovenia}

\title{Enhancing Text2Cypher with Schema Filtering}


\author[1]{Makbule Gulcin Ozsoy}[%
email=makbule.ozsoy@neo4j.com,
]
\cormark[1]
\address[1]{Neo4j, London, UK}
\address[]{Accepted to LLM-TEXT2KG 2025 workshop (in conjunction with ESWC 2025)}


\begin{abstract}
Knowledge graphs represent complex data using nodes, relationships, and properties. Cypher, a powerful query language for graph databases, enables efficient modeling and querying. Recent advancements in large language models allow translation of natural language questions into Cypher queries—Text2Cypher. A common approach is incorporating database schema into prompts. However, complex schemas can introduce noise, increase hallucinations, and raise computational costs. Schema filtering addresses these challenges by including only relevant schema elements, improving query generation while reducing token costs. This work explores various schema filtering methods for Text2Cypher task and analyzes their impact on token length, performance, and cost. Results show that schema filtering effectively optimizes Text2Cypher, especially for smaller models. Consistent with prior research, we find that larger models benefit less from schema filtering due to their longer context capabilities. However, schema filtering remains valuable for both larger and smaller models in cost reduction.

\end{abstract}


\maketitle

\section{Introduction}

Databases are an essential part of modern computer systems for storing and managing data. They are typically accessed via query languages like SQL (for relational databases), SPARQL (for RDF graphs)  or Cypher (for graph databases) which allow users to store and query data for insight \cite{hogan2021knowledge}. Advancements in LLMs have enabled the translation of natural language questions into database queries (Text2SQL, Text2SPARQL, Text2Cypher), allowing non-expert users to query data models on their own terms. 

\begin{figure}
    \centering
    \includegraphics[width=0.6\linewidth]{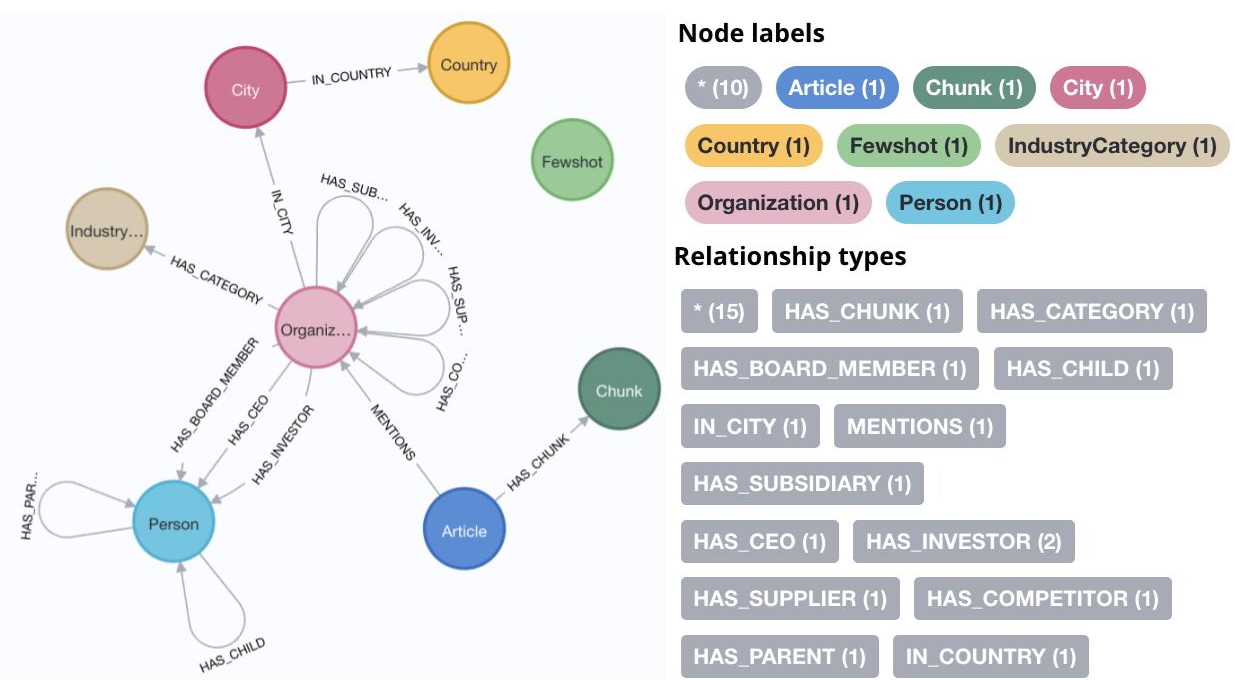} 
    \caption{Overview of an Example Database}
    \label{fig:overall_database}
\end{figure}

To help contextualize an LLM when generating database queries from natural language, a common practice is to incorporate database schema information. Figure \ref{fig:overall_database} shows an example schema where nodes (e.g., Organization, Person) connect through relations (e.g., Has\_CEO, Has\_Investor) with their properties (e.g., name, age). 
Schemas can be provided to LLMs via prompting, but complex schemas introduce noise, increase hallucinations, and raise costs \cite{caferouglu2024sql, chung2025long}. Schema filtering addresses these challenges by selecting only relevant elements, improving query generation while reducing token costs.

In this paper, we apply five schema linking and filtering approaches that improve Text2Cypher: Two static methods that extract the full database schema in different formats and three dynamic methods that prune the schema based on the input question. We evaluate their impact on a Text2Cypher dataset, analyzing token distribution, Cypher generation performance, and cost. Our main contributions are:

\begin{itemize}
    \item We propose new schema filtering techniques. The two static methods use the full database schema in different formats, while our three dynamic methods prune it based on the input question.
    \item We analyze their impact on Text2Cypher task, specifically on prompt token length distribution, query generation performance, and computational cost.
    \item Our results show that schema filtering improves Text2Cypher efficiency. While larger models benefit less due to their extended context windows, smaller models perform better with shorter prompts. Nevertheless, schema filtering remains a cost-effective strategy for all models.
\end{itemize}

The paper is structured as follows: Section \ref{rel_work} covers related work, and Section \ref{schema_filtering} details our schema-filtering approaches for the Text2Cypher task. Section \ref{experiment} presents our experiments and results, and Section \ref{conc} concludes the paper.

\section{Related Work}\label{rel_work}

\subsection{Natural Language to Database Query Language} 

Recent advances in large language models (LLMs) have significantly improved the ability to translate natural language into database query languages.
For instance, there has been extensive research on the Text2SQL and Text2SPARQL tasks, which translates natural language queries to SQL or SPARQL, respectively \cite{yu2018spider, guo2019towards, li2023resdsql, fan2024metasql, shen2024select, lee2025safe, brei2024leveraging, meyer2024assessing, emonet2024llm}. 
Until recently, the Text2Cypher task, which translates natural language into Cypher; the query language used by Neo4j and other graph database systems; had received less attention. However, with advancements in the integration of large language models (LLMs) and knowledge graphs, text-to-graph query language (GQL) tasks, particularly Text2Cypher, have gained increasing interest. 
Several datasets have been developed to support Text2Cypher research, including Opitz and Hochgeschwender \cite{opitz2022zero}, S2CTrans \cite{zhao2023s2ctrans}, CySpider \cite{zhao2023cyspider}, Rel2Graph \cite{zhao2023rel2graph}, SyntheT2C \cite{zhong2024synthet2c}, and Text2Cypher \cite{ozsoy2025text2cypher}. Additionally, studies have explored benchmarking and fine-tuning models for this task, with contributions such as GPT4Graph \cite{guo2023gpt4graph}, TopoChat \cite{xu2024topochat}, Baraki et al. \cite{baraki2024leveraging}, FCAV \cite{liu2150text}, Liang et al. \cite{liang2024aligning} and Text2Cypher \cite{ozsoy2025text2cypher}. In most cases, the baseline model is fine-tuned using prompts that include natural language questions, database schema information, and ground-truth Cypher queries.


\begin{table}
\caption{Instructions used}
  \label{tab:instructions}
  \begin{tabular}{p{0.1\linewidth}p{0.85\linewidth}}
    \hline
    \textbf{Type} & \textbf{Instruction prompt}  \\
    \hline
    System \newline Instruction &  Task: Generate Cypher statement to query a graph database. Instructions: Use only the provided relationship types and properties in the schema. Do not use any other relationship types or properties that are not provided in the schema. Do not include any explanations or apologies in your responses. Do not respond to any questions that might ask anything else than for you to construct a Cypher statement. Do not include any text except the generated Cypher statement. \\
    \hline
    User \newline Instruction & Generate Cypher statement to query a graph database. Use only the provided relationship types and properties in the schema. \newline
            \textbf{Schema: \{schema\}} 
            Question: \{question\} 
            Cypher output: 
         \\
  \hline
\end{tabular}
\end{table}

\begin{figure}
    \centering
    \begin{subfigure}{0.45\textwidth}
        \centering
        \includegraphics[width=\linewidth]{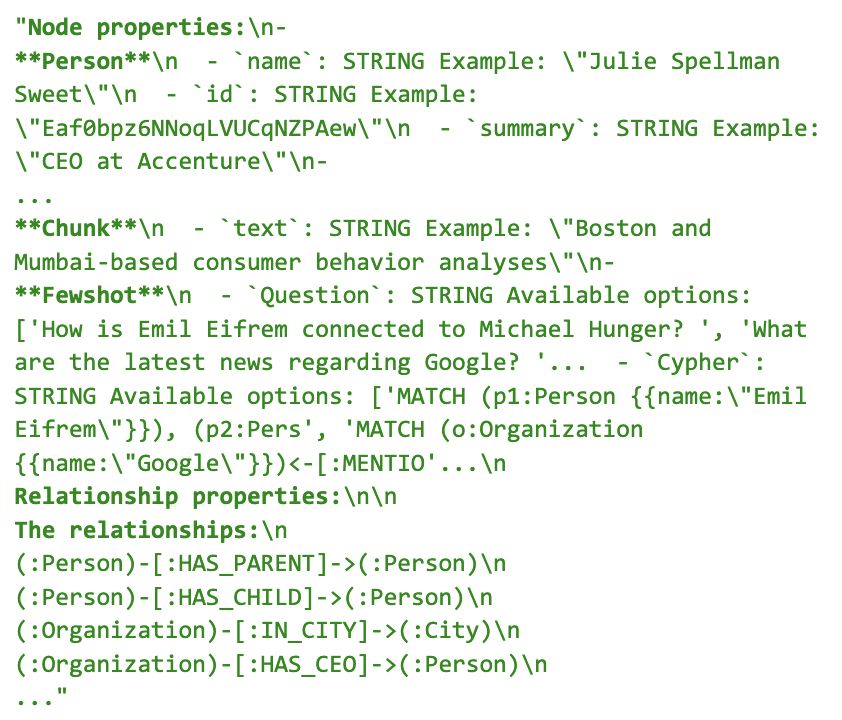}
        \caption{Enhanced Schema}
        \label{fig:enhanced_schema}
    \end{subfigure}
    \hfill
    \begin{subfigure}{0.45\textwidth}
        \centering
        \includegraphics[width=\linewidth]{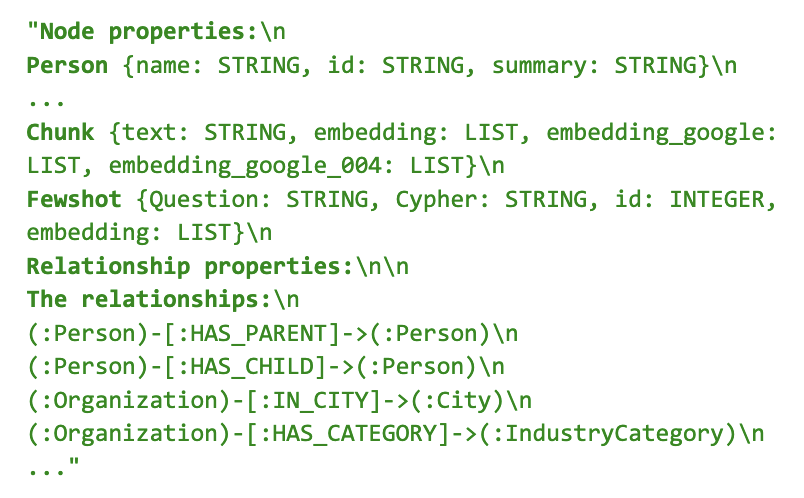}
        \caption{Base Schema}
        \label{fig:base_schema}
    \end{subfigure}

    \caption{Example static schemas for the example database presented in Figure \ref{fig:overall_database}}
    \label{fig:static_schema}
\end{figure}

\subsection{Schema Filtering in Query Generation}
Schema information is essential for accurate query generation, ensuring correct linking of query terms to database structures \cite{caferouglu2024sql, chung2025long}. This process, known as schema linking, plays a key role in Text2SQL and Text2Cypher tasks by mapping query words to relevant database elements \cite{lei2020re, caferouglu2024sql}. 
While providing the full schema in the prompt is possible, schema filtering is often preferred to reduce noise, computational cost, and hallucinations \cite{caferouglu2024sql, cao2024rsl}. However, we must remain aware that excessive filtering can remove essential components, harming accuracy \cite{maamari2024death}. 
Early Text2SQL schema filtering relied on heuristics like string matching, as seen in IRNet \cite{guo2019towards} and TypeSQL \cite{yu2018typesql}. Later, learning-based methods such as Dong et al. \cite{dong2019data}, Bogin et al. \cite{bogin2019global}, and RAT-SQL \cite{wang2019rat} were proposed. Recent approaches utilize LLMs through prompting, fine-tuning, or agent-based techniques, such as DIN-SQL \cite{pourreza2023din}, RESDSQL \cite{li2023resdsql}, CHESS \cite{talaei2024chess}, E-SQL \cite{caferouglu2024sql}, ExSL \cite{glass2025extractive}and KaSLA \cite{yuan2025knapsack}. 
While schema filtering is common, studies suggest it is less necessary for LLMs with long context windows but remains valuable for smaller models \cite{maamari2024death, caferouglu2024sql, chung2025long}. The trade-off is, however,  that larger context sizes increase latency and computational cost for complex databases, making filtering highly beneficial \cite{chung2025long}. 

Research on schema filtering for Text2Cypher or other graph query languages (Text2GQL) is presently limited compared to Text2SQL. Liang et al. \cite{liang2024aligning} explored aligning LLMs for a Text2GQL task in Chinese, using a schema filtering module that executes: (i) extraction of the database schema as a dictionary, (ii) extraction of the named entities from the query, and (iii) mapping these entities to the schema dictionary. For queries requiring multiple nodes and relations, they used A* algorithm \cite{hart1968formal} to find the shortest path. 
NAT-NL2GQL \cite{liang2024nat} includes a module for preprocessing inputs and executing schema extraction, following a similar approach to Liang et al. \cite{liang2024aligning}. Additionally, they use an LLM for filtering multiple matched schema items before proceeding with the Text2GQL task. 
In this work, we examine the impact of schema filtering on the Text2Cypher task, focusing on both performance and cost. 

\section{Schema Filtering for Text2Cypher}\label{schema_filtering}
We now present schema filtering for Text2Cypher using a template from \cite{ozsoy2025text2cypher} (Table \ref{tab:instructions}), focusing on the schema field with two static and three dynamic formats.

\begin{figure}
    \centering
    \begin{minipage}[b]{0.45\textwidth}
        \begin{subfigure}{\linewidth}
            \includegraphics[width=0.9\linewidth]{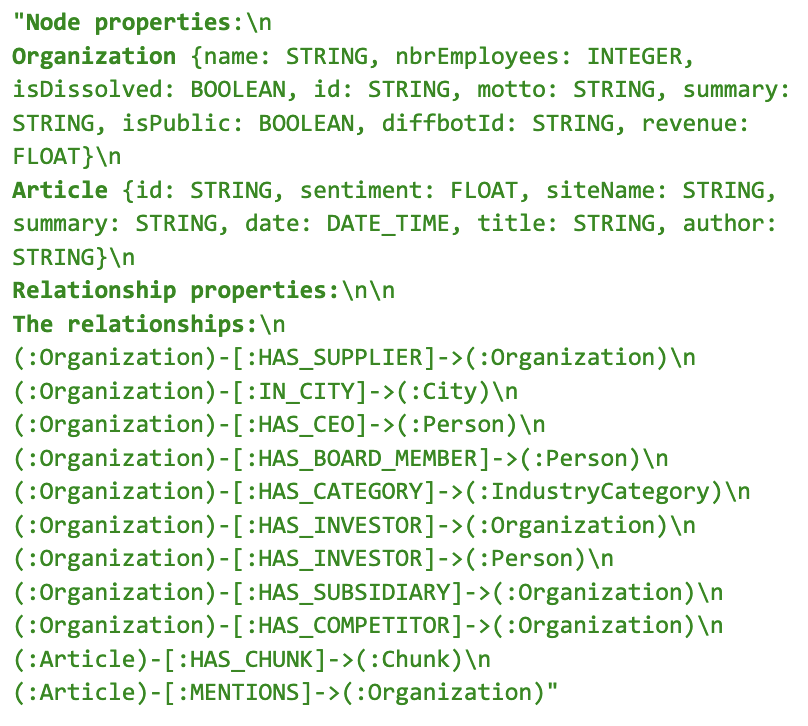}
            \caption{Pruned By Exact-Match Schema}
            \label{fig:pruned_by_exact_match_schema}
        \end{subfigure}

        \vspace{1em} 

        \begin{subfigure}{\linewidth}
            \includegraphics[width=0.9\linewidth]{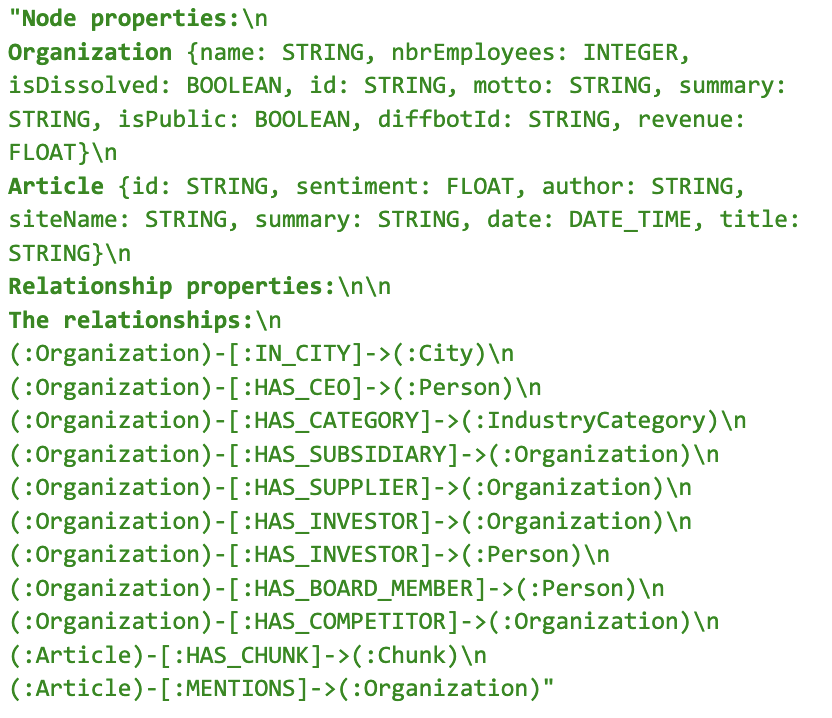}
            \caption{NER Masked \& Pruned by Exact-Match Schema}
            \label{fig:ner_masked_pruned_by_exact_match_schema}
        \end{subfigure}
    \end{minipage}
    \hfill
    \begin{minipage}[b]{0.45\textwidth}
        \begin{subfigure}{\linewidth}
            \includegraphics[width=0.9\linewidth]{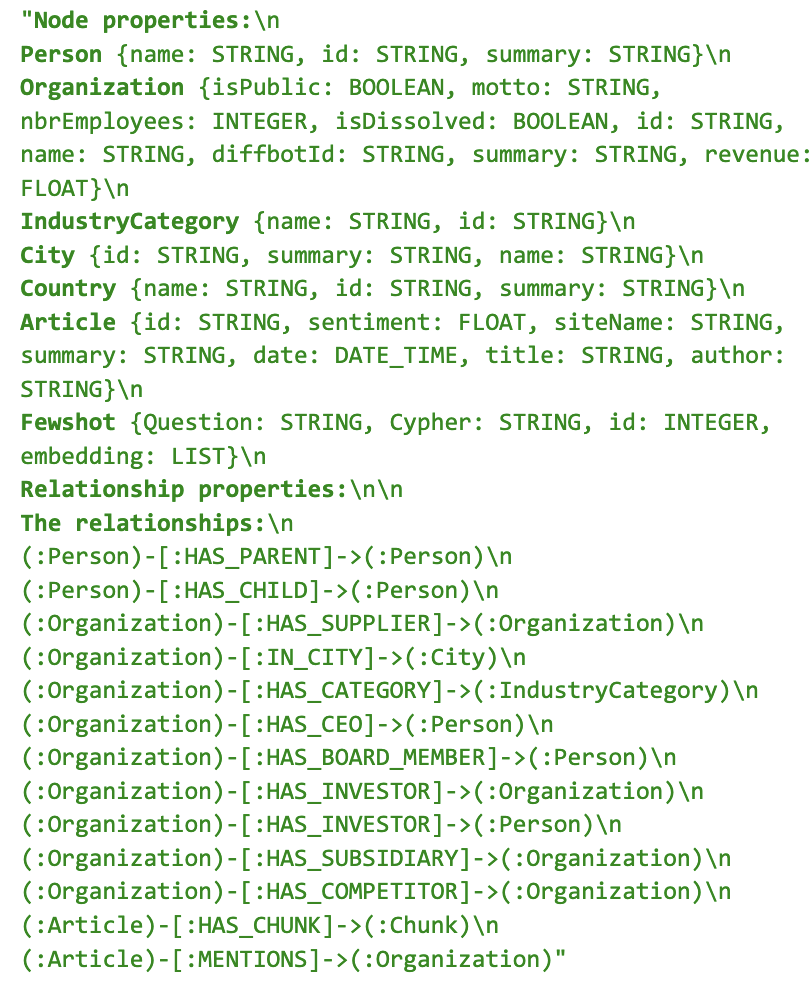}
            \caption{Pruned by Similarity Schema}
            \label{fig:pruned_by_similarity_schema}
        \end{subfigure}
    \end{minipage}

    \caption{Dynamic schemas for the question: "List the articles that mention the organization 'Acme Group'."}
    \label{fig:dynamic_schema}
\end{figure}

\subsection{Static Schemas}
Cypher is the query language for Neo4j, a graph database. Neo4j offers various tools for retrieving database schema information, based on the database structure rather than the input query. While this allows efficient caching, it leads to longer schema representations, increasing token length and context requirements for LLMs. 
We utilized two static schema formats provided by Neo4j frameworks:  
\begin{itemize}
    \item \textbf{Enhanced Schema}: This is one of the default schema types provided by Neo4j. It provides an enhanced view of the database schema, including list of nodes, relationships and their properties. It additionally  provides example values for the fields.  For instance, if the property is the 'name' of the 'Actor' node, examples might include: ['Tom Hanks', 'Julia Roberts', ...]. An example enhanced schema is presented in Figure \ref{fig:enhanced_schema}.

    \item \textbf{Base Schema}: This is another default schema types provided by Neo4j. It provides similar information as the Enhanced Schema, except it does not include examples of properties, and the formatting is different. An example for this schema format is presented in Figure \ref{fig:base_schema}. 
\end{itemize}

\subsection{Dynamically Pruned Schemas}
We implement three dynamic schema filtering approaches, which prune the baseline schemas based on the input natural language question.

\begin{itemize}
    \item \textbf{Pruned By Exact-Match}: This approach compares node labels, relationship types, and properties to words in the input question. Similar to Liang et al. \cite{liang2024aligning} and NAT-NL2GQL \cite{liang2024nat}, if an exact case-insensitive match is found, the corresponding schema elements are retained; otherwise, they are removed. Our method also considers properties as well as labels, and we retain multiple matching elements (e.g., synonyms) to prevent excessive pruning. See Figure \ref{fig:pruned_by_exact_match_schema} for an example.

    \item \textbf{NER Masked \& Pruned By Exact-Match}: This approach replaces named entities with their entity types before applying exact-match filtering. NER-masking prevents irrelevant matches. For example, in the query "List the articles that mention the organization 'Acme Energy'," it avoids incorrect matches, such as retaining properties of a node labeled 'Energy,' which is unrelated. See Figure \ref{fig:ner_masked_pruned_by_exact_match_schema} for an example.
    
    \item \textbf{Pruned by Similarity}: This approach extends exact-match pruning by incorporating similarity-based filtering. Instead of requiring an exact match, it computes similarity scores between query terms and schema elements, retaining only those above a predefined threshold. While various similarity measures could be used, we rely on embedding-based similarity. An example of this schema filtering approach is shown in Figure \ref{fig:pruned_by_similarity_schema}.
\end{itemize}

\section{Experimental Setup and Results }\label{experiment}

\subsection{Experimental Setup and Evaluation Metrics}
We conducted experiments using a publicly available Text2Cypher dataset \cite{ozsoy2025text2cypher}, focusing on a subset with accessible databases for query execution, resulting in 22,093 training and 2,471 test samples. Schema filtering was assessed using the 'unsloth/Meta-Llama-3.1-8B-Instruct-bnb-4bit', ' unsloth/Qwen2.5-7B-Instruct-bnb-4bit' and 'GoogleAIStudio/Gemini-1.5-Flash' models, referred as Llama-3.1-8B, Qwen2.5-7B and Gemini-1.5-Flash, respectively, in the remainder of the paper.
For Cypher generation, after utilizing the LLMs, an additional post-processing step is executed to remove unwanted text, such as 'cypher:' suffix. Furthermore, the spaCy framework is used for named entity extraction and similarity computations. 
To compute evaluation metrics, we used the Hugging Face Evaluate library \cite{hfEvaluate}. We employed two evaluation procedures: 
(i) Translation-based (Lexical) evaluation: Compares generated Cypher queries with reference queries based on text content. We used Google-BLEU score while presenting the results.
(ii) Execution-based evaluation: Executes both generated and reference queries on target databases and compares their outputs (sorted lexicographically) using the same metrics as the translation-based evaluation. We used ExactMatch score while presenting the results.

\subsection{Evaluation Results}
We evaluate the proposed schema formats based on (i) token distribution and cost, and (ii) performance. 

\begin{figure}
    \centering

    \begin{subfigure}{0.45\textwidth}
        \centering
        \includegraphics[width=\linewidth]{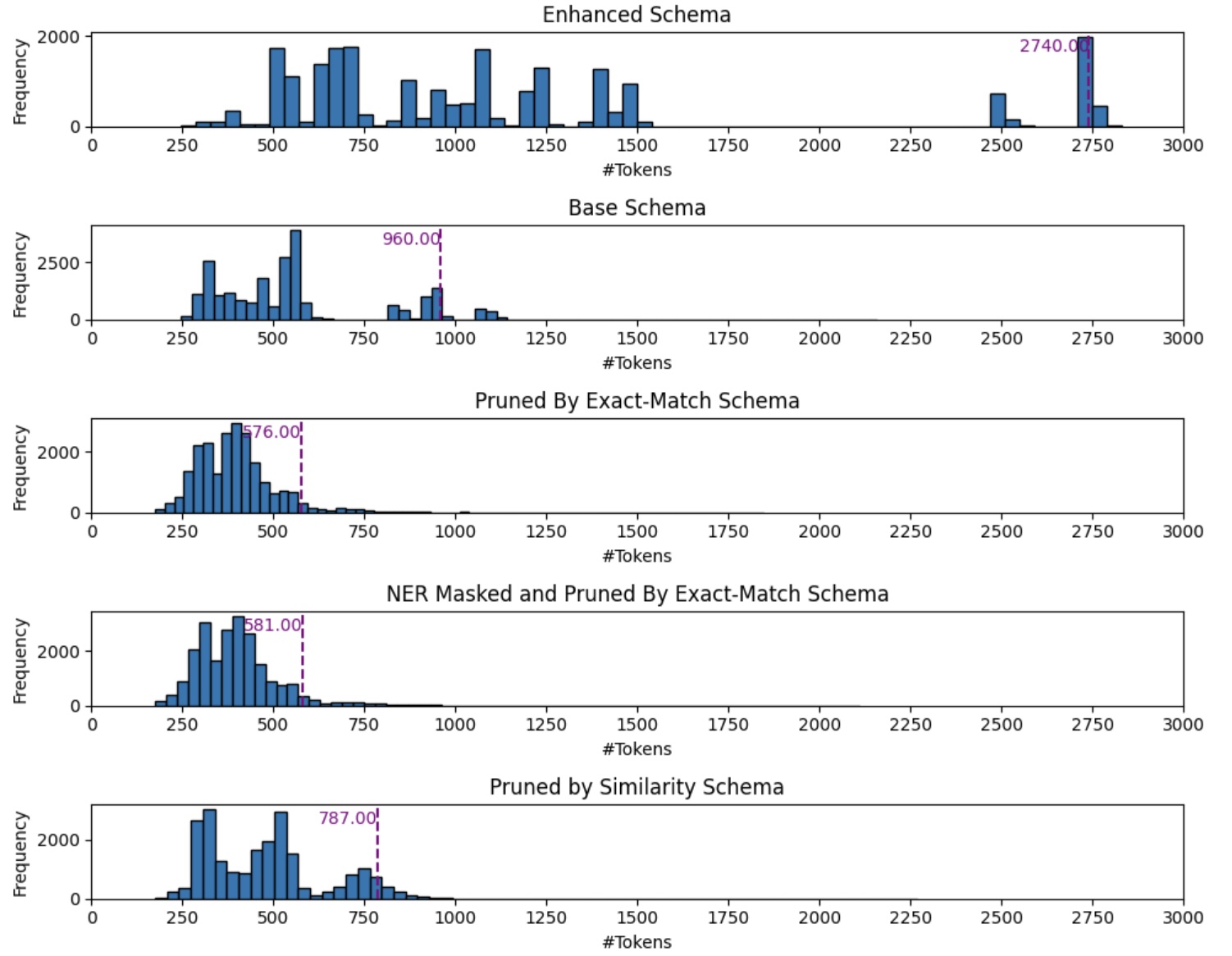}
        \caption{Token Distribution on Training Set}
        \label{fig:train_token_distr}
    \end{subfigure}
    \hfill
    \begin{subfigure}{0.45\textwidth}
        \centering
        \includegraphics[width=\linewidth]{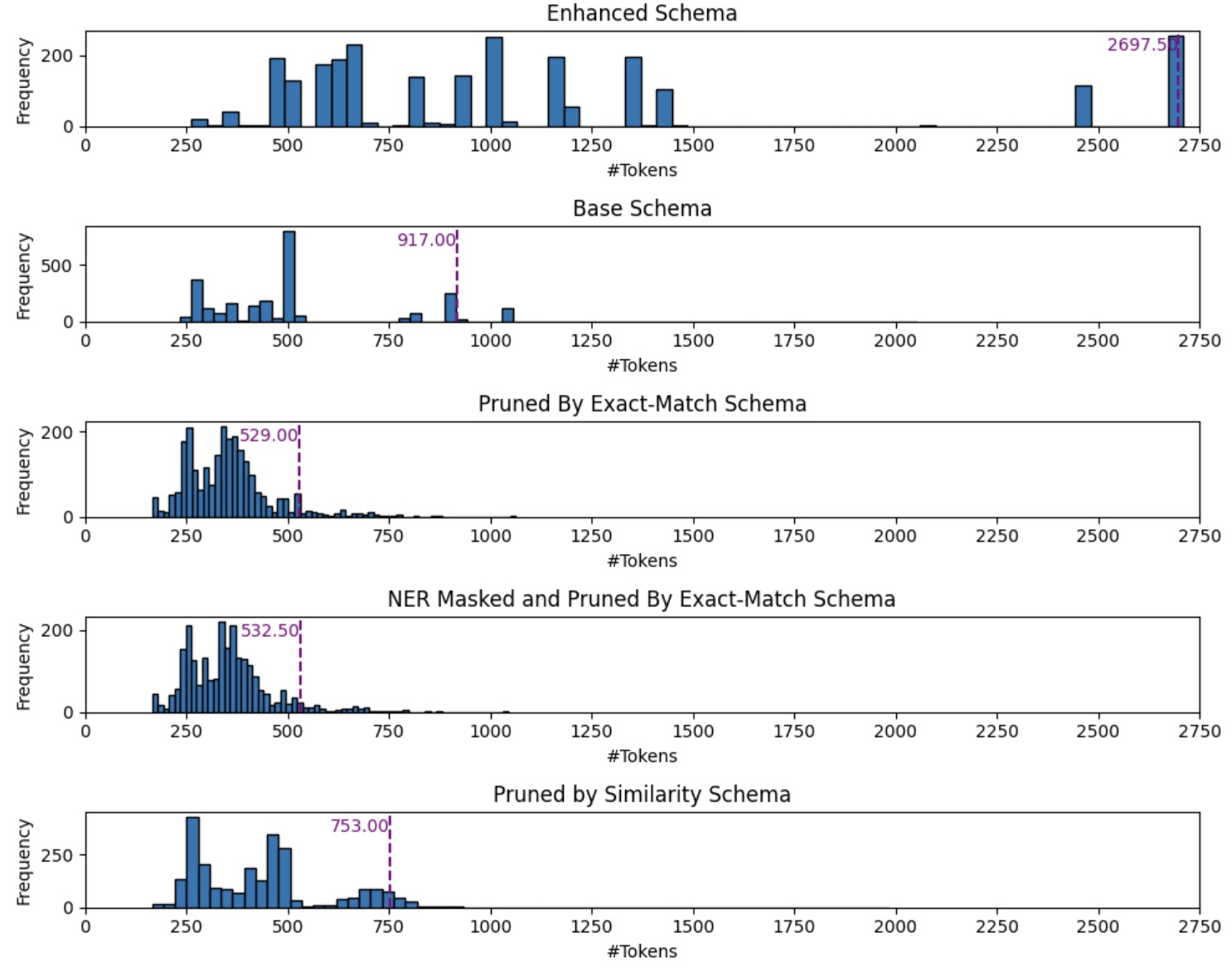}
        \caption{Token Distribution on Test Set}
        \label{fig:test_token_distr}
    \end{subfigure}

    \caption{Token distributions of the training and test sets, based on the tokenizer from the Llama-3.1-8B model. The 95th percentile (p95) value is marked with a purple line.}
    \label{fig:token_distr}
\end{figure}

\begin{table}
  \centering  
  \begin{tabular}{p{0.25\linewidth} p{0.08\linewidth} p{0.08\linewidth} p{0.08\linewidth} p{0.08\linewidth}}
      \hline
      \textbf{Schema Type} & \textbf{min} & \textbf{max} & \textbf{median} & \textbf{p95} \\
      \hline
      \textbf{Enhanced Schema} & 263 & 2710 & 921 & 2697 \\
      \textbf{Base Schema} & 233 & 2052 & 494 & 917 \\
      \textbf{Pruned By Exact-Match} & 166 & 1063 & 344 & 529 \\
      \textbf{NER Masked \& Pruned By Exact-Match} & 166 & 1045 & 344 & 532 \\
      \textbf{Pruned by Similarity} & 166 & 1982 & 422 & 753 \\
      \hline
  \end{tabular}
  \caption{Token distribution statistics of the test set}
  \label{tab:token_distr_stats}
\end{table}

\subsubsection{Impact on Token Distribution \& Cost}
Schema format impacts both prompt length and token count. For example, with the Llama-3.1-8B tokenizer, the base prompt is about 150 tokens, but adding schema information increases it to over 2,700 tokens. Figure \ref{fig:token_distr} shows token distributions for training and test sets. Table \ref{tab:token_distr_stats} provides additional token details for the test set. 
Results show that the Enhanced Schema leads to the longest prompts, while switching to the Base Schema reduces the P95 token length by one-third. Exact-match pruning (with or without NER masking) further reduces the P95 token length to 1/6th of the original. Similarity-based pruning increases schema length but reduces the P95 token length to about 1/4th of the original.

Reducing the token count reduces costs, whether for LLM vendor payments or infrastructure expenses for self-hosted models (e.g., storage and GPU access). In a scenario with 20,000 instances, where input token length aligns with the median in Table \ref{tab:token_distr_stats}, we compare costs across models (Table \ref{tab:cost}). In the table, we assume output lengths remain constant and only input tokens contribute to the cost. The results show that cost scales linearly with token usage, but factors like output token count, caching, and batch processing can affect this. Shorter prompts lead to significant cost reductions.

While dynamic pruning reduces token length and costs, it may introduce computational overhead as a side-effect. Unlike Enhanced or Base Schema (which are cached), dynamic pruning is performed for each query, which might increase latency. However, we observe this overhead is minimal, especially for methods like ‘Pruned by Exact-Match,’ which uses regular expression matching.

\begin{table}
\caption{Schema type vs. Cost. Notes: Median token counts: Table \ref{tab:token_distr_stats}, Number of instances: 20K, Prices-Feb. 27, 2025: (i) GoogleAIStudio/Gemini-2.0-Flash: \$0.15 / 1M tokens (ii) Anthropic/Claude 3.5 Haiku: \$0.80 / 1M tokens (iii) Self hosted/LLaMA-3.1-8B: Assumed A40 48GB machine on RunPod (\$0.44 per hour) with 20 tokens/sec}
  \label{tab:cost}
  \begin{tabular}{P{0.25\linewidth}cP{0.13\linewidth}P{0.13\linewidth}P{0.15\linewidth}}
    \hline
    \textbf{Schema Type}  & \textbf{\# Total tokens} & \textbf{Gemini-2.0 \newline Flash} & \textbf{Claude-3.5 \newline Haiku} &\textbf{LLaMA-3.1-8B \newline (Self Hosted)}\\
    \hline
    \textbf{Enhanced Schema} & $20K * 921$ \newline $=\sim$18.5M & \$2.7 & \$14.7 & \$112.5 \\
    \textbf{Base Schema} & $20K * 494$ \newline $=\sim$10M  & \$1.5 & \$7.9 &\$60.3 \\
    \textbf{Pruned By Exact-Match}  & $20K * 344$ \newline $=\sim$7M  & \$1.0  & \$5.5 & \$42.0\\
    \textbf{NER Masked \& \newline Pruned By Exact-Match}  & $20K * 344$ \newline $=\sim$7M  & \$1.0 & \$5.5  & \$42.0\\
    \textbf{Pruned by Similarity} & $20K * 422$ \newline $=\sim$8.5M  & \$1.3 & \$6.7  &  \$51.5\\
    
\hline
\end{tabular}
\end{table}

\subsubsection{Impact on Performance}
\begin{figure}
    \centering
    \begin{subfigure}[b]{0.25\textwidth}
        \includegraphics[width=0.98\linewidth]{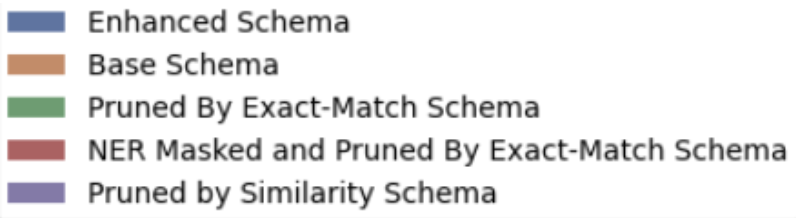}
        \caption{Schema formats}
        \label{fig:performance_labels}
    \end{subfigure}
    \hfill
    \begin{subfigure}[b]{0.37\textwidth}
        \includegraphics[width=\linewidth]{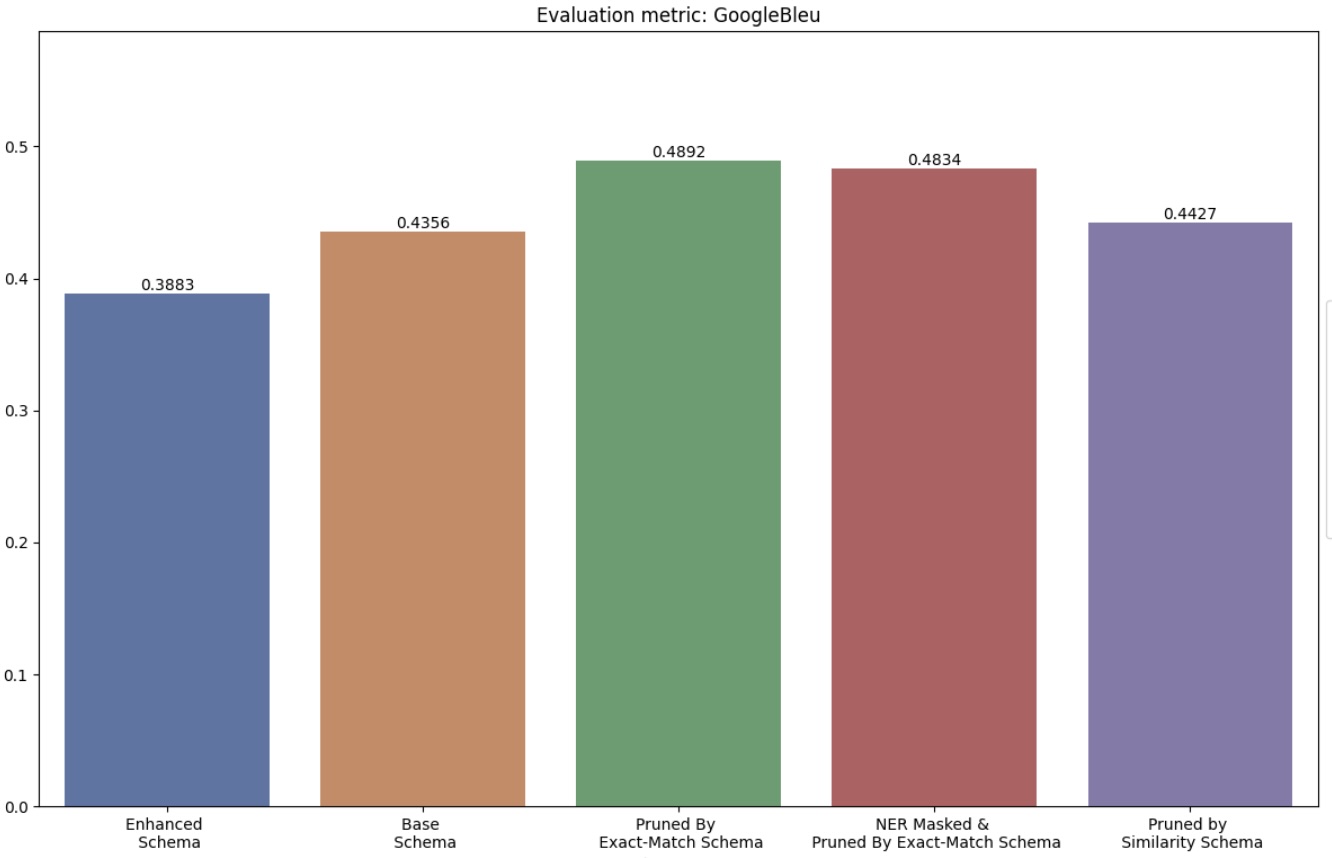}
        \caption{Translation-based - Google-Bleu score}
        \label{fig:llama_google_blue_score}
    \end{subfigure}
    \hfill
    \begin{subfigure}[b]{0.37\textwidth}
        \includegraphics[width=\linewidth]{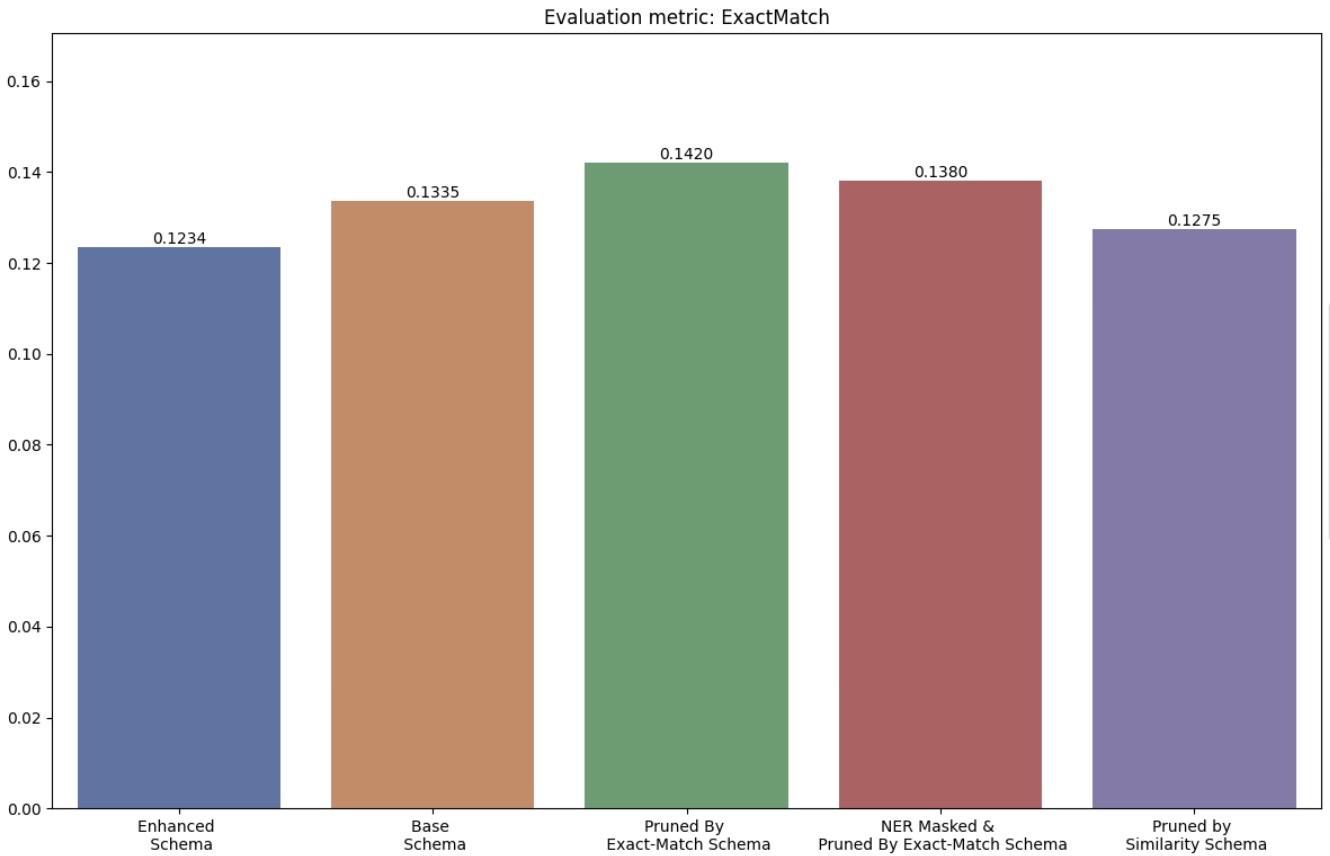}
        \caption{Execution-based - Exact-match score}
        \label{fig:llama_exact_match_score}
    \end{subfigure}
    \caption{Performance across various schema formats}
    \label{fig:llama_result}
\end{figure}

We evaluate the impact of proposed schema formats on Text2Cypher performance using the Llama-3.1-8B model. Figure \ref{fig:llama_result} presents the results, showing that longer prompts lead to lower performance. The highest accuracy is achieved with the ‘Pruned by Exact-Match Schema.’ NER masking and Similarity-Based Matching did not improve performance but may be beneficial for other datasets.

\begin{figure}
    \centering
    \begin{subfigure}[b]{0.45\textwidth}
        \includegraphics[width=0.9\linewidth]{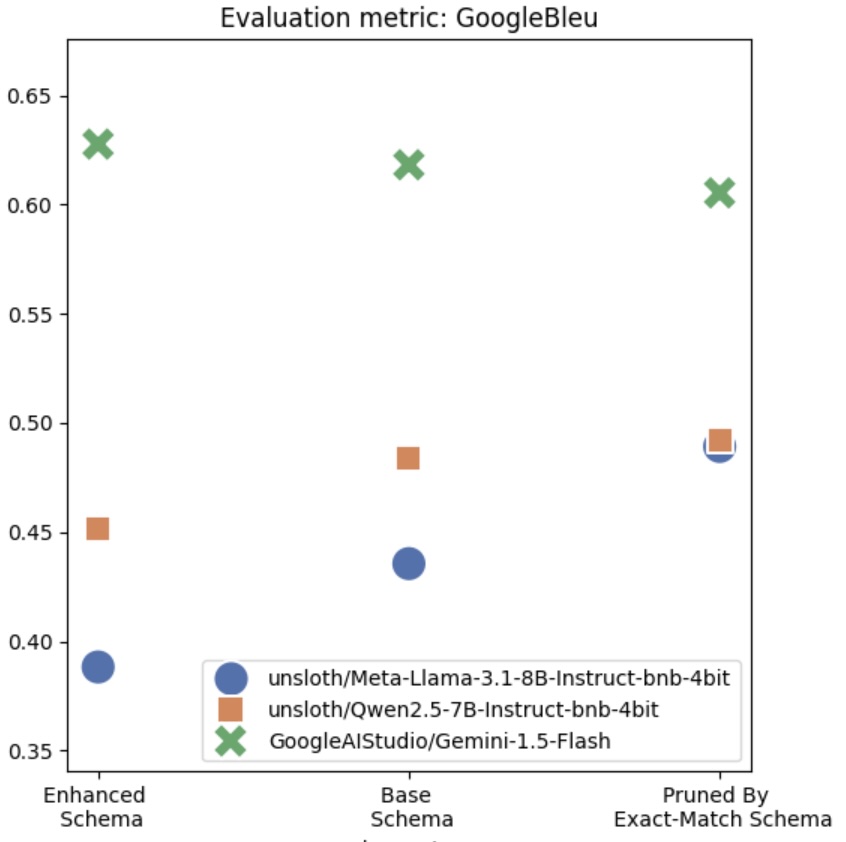}
        \caption{Translation-based - Google-Bleu score}
        \label{fig:compare_google_blue_score}
    \end{subfigure}
    \hfill
    \begin{subfigure}[b]{0.45\textwidth}
        \includegraphics[width=0.9\linewidth]{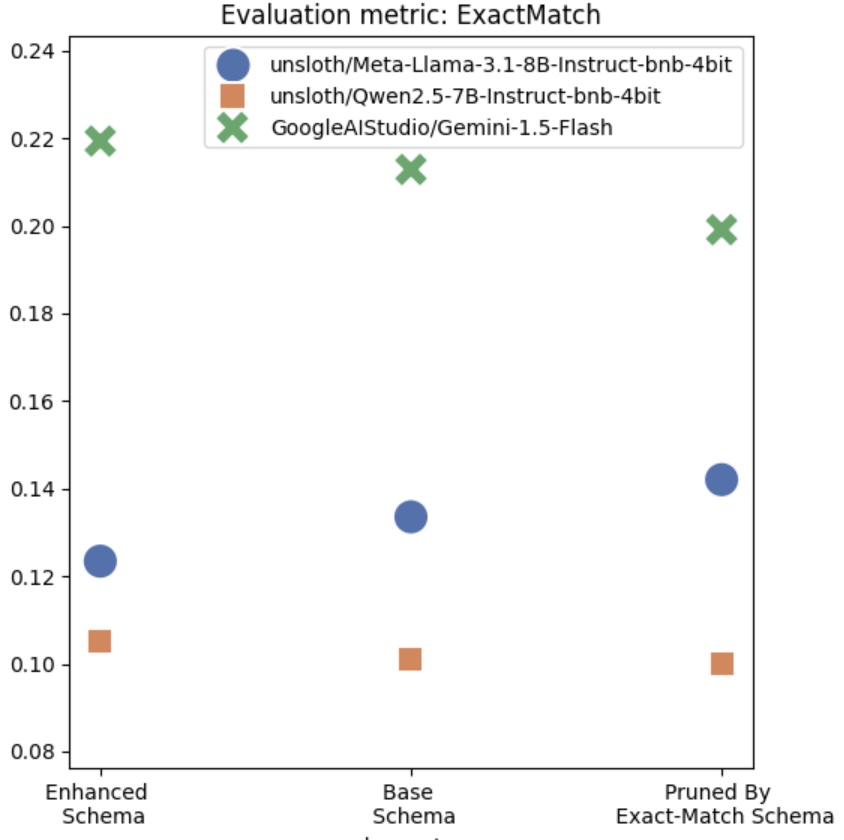}
        \caption{Execution-based - Exact-match score}
        \label{fig:compare_exact_match_score}
    \end{subfigure}
    \caption{Performance comparison of various models}
    \label{fig:compare_results}
\end{figure}

We further compared the performance of different LLMs on a selected subset of schema formats. In addition to Llama-3.1-8B, we evaluated Qwen2.5-7B and Gemini-1.5-Flash. While Llama-3.1-8B and Qwen2.5-7B are similar in size, they differ in multiple ways, such as tokenization strategies. 
Gemini-1.5-Flash, in contrast, has a larger model size and a significantly longer context window. For comparison, we used three schema formats—Enhanced, Base, and Pruned by Exact-Match—with decreasing token lengths. 
Figure \ref{fig:compare_results} presents the results, highlighting key trends:
(i) In terms of lexical (translation-based) comparison, performance of Llama-3.1-8B and Qwen2.5-7B models are improved as prompt length decreased. However, Gemini-1.5-Flash had the opposite trend, performing better with longer prompts. The drop in Gemini-1.5-Flash for shorter prompts was minor, remaining below 5\%. 
(ii) In terms of execution-based evaluation, Llama-3.1-8B model showed improved performance with shorter prompts, while Qwen2.5-7B and Gemini-1.5-Flash experienced slight declines, both around 2\%. 
These findings align with observation made by previous research \cite{caferouglu2024sql, chung2025long}: The impact of schema length varies across models, with Gemini-1.5-Flash potentially benefiting from longer context while the other smaller models perform better with shorter inputs.



\section{Conclusion} \label{conc}
We presented schema filtering for Text2Cypher and analyzed its effects on token length, performance, and cost. 
We found that reducing schema size improved performance for most models, and reduced cost for all of those we tested. 
Comparison of various models revealed that smaller models performed better with shorter prompts, while larger models benefited from longer contexts. Dynamically pruning schemas reduced both token counts and cost, introducing slightly more latency but remained the most efficient overall. 
This work has two main limitations. First, experiments used a subset of a public dataset \cite{ozsoy2025text2cypher}, selecting instances with accessible databases for schema extraction. However, these demo-oriented databases often have simpler schemas than real-world ones. 
The longest schema had around 2700 tokens, whereas real databases likely have more complex structures, making schema filtering more critical. 
Second, our filtering methods are heuristic-based. More advanced techniques, like those in Text2SQL (see Section \ref{rel_work}), may yield better results and require further exploration. 
In the future, we will explore adaptive schema selection based on model characteristics, as well as the impact of schema filtering on the fine-tuning process and its effects on fine-tuned models. 


\section*{Declaration on Generative AI}

  
During the preparation of this work, the author(s) used Chat-GPT in order to: 'Improve writing style' and 'Paraphrase and reword'. After using these tool(s)/service(s), the author(s) reviewed and edited the content as needed and take(s) full responsibility for the publication’s content. 

\bibliography{main-arxiv}

\begin{thebibliography}{37}
\expandafter\ifx\csname natexlab\endcsname\relax\def\natexlab#1{#1}\fi
\providecommand{\url}[1]{\texttt{#1}}
\providecommand{\href}[2]{#2}
\providecommand{\path}[1]{#1}
\providecommand{\DOIprefix}{doi:}
\providecommand{\ArXivprefix}{arXiv:}
\providecommand{\URLprefix}{URL: }
\providecommand{\Pubmedprefix}{pmid:}
\providecommand{\doi}[1]{\href{http://dx.doi.org/#1}{\path{#1}}}
\providecommand{\Pubmed}[1]{\href{pmid:#1}{\path{#1}}}
\providecommand{\bibinfo}[2]{#2}
\ifx\xfnm\relax \def\xfnm[#1]{\unskip,\space#1}\fi
\bibitem[{Hogan et~al.(2021)Hogan, Blomqvist, Cochez, d’Amato, Melo, Gutierrez, Kirrane, Gayo, Navigli, Neumaier et~al.}]{hogan2021knowledge}
\bibinfo{author}{A.~Hogan}, \bibinfo{author}{E.~Blomqvist}, \bibinfo{author}{M.~Cochez}, \bibinfo{author}{C.~d’Amato}, \bibinfo{author}{G.~D. Melo}, \bibinfo{author}{C.~Gutierrez}, \bibinfo{author}{S.~Kirrane}, \bibinfo{author}{J.~E.~L. Gayo}, \bibinfo{author}{R.~Navigli}, \bibinfo{author}{S.~Neumaier}, et~al.,
\newblock \bibinfo{title}{Knowledge graphs},
\newblock \bibinfo{journal}{ACM Computing Surveys (Csur)} \bibinfo{volume}{54} (\bibinfo{year}{2021}) \bibinfo{pages}{1--37}.
\bibitem[{Cafero{\u{g}}lu and Ulusoy(2024)}]{caferouglu2024sql}
\bibinfo{author}{H.~A. Cafero{\u{g}}lu}, \bibinfo{author}{{\"O}.~Ulusoy},
\newblock \bibinfo{title}{E-sql: Direct schema linking via question enrichment in text-to-sql},
\newblock \bibinfo{journal}{arXiv preprint arXiv:2409.16751}  (\bibinfo{year}{2024}).
\bibitem[{Chung et~al.(2025)Chung, Kakkar, Gan, Milne, and Ozcan}]{chung2025long}
\bibinfo{author}{Y.~Chung}, \bibinfo{author}{G.~T. Kakkar}, \bibinfo{author}{Y.~Gan}, \bibinfo{author}{B.~Milne}, \bibinfo{author}{F.~Ozcan},
\newblock \bibinfo{title}{Is long context all you need? leveraging llm's extended context for nl2sql},
\newblock \bibinfo{journal}{arXiv preprint arXiv:2501.12372}  (\bibinfo{year}{2025}).
\bibitem[{Yu et~al.(2018)Yu, Zhang, Yang, Yasunaga, Wang, Li, Ma, Li, Yao, Roman et~al.}]{yu2018spider}
\bibinfo{author}{T.~Yu}, \bibinfo{author}{R.~Zhang}, \bibinfo{author}{K.~Yang}, \bibinfo{author}{M.~Yasunaga}, \bibinfo{author}{D.~Wang}, \bibinfo{author}{Z.~Li}, \bibinfo{author}{J.~Ma}, \bibinfo{author}{I.~Li}, \bibinfo{author}{Q.~Yao}, \bibinfo{author}{S.~Roman}, et~al.,
\newblock \bibinfo{title}{Spider: A large-scale human-labeled dataset for complex and cross-domain semantic parsing and text-to-sql task},
\newblock \bibinfo{journal}{arXiv preprint arXiv:1809.08887}  (\bibinfo{year}{2018}).
\bibitem[{Guo et~al.(2019)Guo, Zhan, Gao, Xiao, Lou, Liu, and Zhang}]{guo2019towards}
\bibinfo{author}{J.~Guo}, \bibinfo{author}{Z.~Zhan}, \bibinfo{author}{Y.~Gao}, \bibinfo{author}{Y.~Xiao}, \bibinfo{author}{J.-G. Lou}, \bibinfo{author}{T.~Liu}, \bibinfo{author}{D.~Zhang},
\newblock \bibinfo{title}{Towards complex text-to-sql in cross-domain database with intermediate representation},
\newblock \bibinfo{journal}{arXiv preprint arXiv:1905.08205}  (\bibinfo{year}{2019}).
\bibitem[{Li et~al.(2023)Li, Zhang, Li, and Chen}]{li2023resdsql}
\bibinfo{author}{H.~Li}, \bibinfo{author}{J.~Zhang}, \bibinfo{author}{C.~Li}, \bibinfo{author}{H.~Chen},
\newblock \bibinfo{title}{Resdsql: Decoupling schema linking and skeleton parsing for text-to-sql},
\newblock in: \bibinfo{booktitle}{Proceedings of the AAAI Conference on Artificial Intelligence}, volume~\bibinfo{volume}{37}, \bibinfo{year}{2023}, pp. \bibinfo{pages}{13067--13075}.
\bibitem[{Fan et~al.(2024)Fan, He, Ren, Huang, Jing, Zhang, and Wang}]{fan2024metasql}
\bibinfo{author}{Y.~Fan}, \bibinfo{author}{Z.~He}, \bibinfo{author}{T.~Ren}, \bibinfo{author}{C.~Huang}, \bibinfo{author}{Y.~Jing}, \bibinfo{author}{K.~Zhang}, \bibinfo{author}{X.~S. Wang},
\newblock \bibinfo{title}{Metasql: A generate-then-rank framework for natural language to sql translation},
\newblock \bibinfo{journal}{arXiv preprint arXiv:2402.17144}  (\bibinfo{year}{2024}).
\bibitem[{Shen and Kejriwal(2024)}]{shen2024select}
\bibinfo{author}{K.~Shen}, \bibinfo{author}{M.~Kejriwal},
\newblock \bibinfo{title}{Select-sql: Self-correcting ensemble chain-of-thought for text-to-sql},
\newblock \bibinfo{journal}{arXiv preprint arXiv:2409.10007}  (\bibinfo{year}{2024}).
\bibitem[{Lee et~al.(2025)Lee, Baek, Kim, and Lee}]{lee2025safe}
\bibinfo{author}{J.~Lee}, \bibinfo{author}{I.~Baek}, \bibinfo{author}{B.~Kim}, \bibinfo{author}{H.~Lee},
\newblock \bibinfo{title}{Safe-sql: Self-augmented in-context learning with fine-grained example selection for text-to-sql},
\newblock \bibinfo{journal}{arXiv preprint arXiv:2502.11438}  (\bibinfo{year}{2025}).
\bibitem[{Brei et~al.(2024)Brei, Frey, and Meyer}]{brei2024leveraging}
\bibinfo{author}{F.~Brei}, \bibinfo{author}{J.~Frey}, \bibinfo{author}{L.-P. Meyer},
\newblock \bibinfo{title}{Leveraging small language models for text2sparql tasks to improve the resilience of ai assistance},
\newblock \bibinfo{journal}{arXiv preprint arXiv:2405.17076}  (\bibinfo{year}{2024}).
\bibitem[{Meyer et~al.(2024)Meyer, Frey, Brei, and Arndt}]{meyer2024assessing}
\bibinfo{author}{L.-P. Meyer}, \bibinfo{author}{J.~Frey}, \bibinfo{author}{F.~Brei}, \bibinfo{author}{N.~Arndt},
\newblock \bibinfo{title}{Assessing sparql capabilities of large language models},
\newblock \bibinfo{journal}{arXiv preprint arXiv:2409.05925}  (\bibinfo{year}{2024}).
\bibitem[{Emonet et~al.(2024)Emonet, Bolleman, Duvaud, de~Farias, and Sima}]{emonet2024llm}
\bibinfo{author}{V.~Emonet}, \bibinfo{author}{J.~Bolleman}, \bibinfo{author}{S.~Duvaud}, \bibinfo{author}{T.~M. de~Farias}, \bibinfo{author}{A.~C. Sima},
\newblock \bibinfo{title}{Llm-based sparql query generation from natural language over federated knowledge graphs},
\newblock \bibinfo{journal}{arXiv preprint arXiv:2410.06062}  (\bibinfo{year}{2024}).
\bibitem[{Opitz and Hochgeschwender(2022)}]{opitz2022zero}
\bibinfo{author}{D.~Opitz}, \bibinfo{author}{N.~Hochgeschwender},
\newblock \bibinfo{title}{From zero to hero: generating training data for question-to-cypher models},
\newblock in: \bibinfo{booktitle}{Proceedings of the 1st International Workshop on Natural Language-based Software Engineering}, \bibinfo{year}{2022}, pp. \bibinfo{pages}{17--20}.
\bibitem[{Zhao et~al.(2023{\natexlab{a}})Zhao, Ge, Shen, Hu, and Wang}]{zhao2023s2ctrans}
\bibinfo{author}{Z.~Zhao}, \bibinfo{author}{X.~Ge}, \bibinfo{author}{Z.~Shen}, \bibinfo{author}{C.~Hu}, \bibinfo{author}{H.~Wang},
\newblock \bibinfo{title}{S2ctrans: Building a bridge from sparql to cypher},
\newblock in: \bibinfo{booktitle}{International Conference on Database and Expert Systems Applications}, \bibinfo{organization}{Springer}, \bibinfo{year}{2023}{\natexlab{a}}, pp. \bibinfo{pages}{424--430}.
\bibitem[{Zhao et~al.(2023{\natexlab{b}})Zhao, Liu, French, and Stewart}]{zhao2023cyspider}
\bibinfo{author}{Z.~Zhao}, \bibinfo{author}{W.~Liu}, \bibinfo{author}{T.~French}, \bibinfo{author}{M.~Stewart},
\newblock \bibinfo{title}{Cyspider: A neural semantic parsing corpus with baseline models for property graphs},
\newblock in: \bibinfo{booktitle}{Australasian Joint Conference on Artificial Intelligence}, \bibinfo{organization}{Springer}, \bibinfo{year}{2023}{\natexlab{b}}, pp. \bibinfo{pages}{120--132}.
\bibitem[{Zhao et~al.(2023{\natexlab{c}})Zhao, Liu, French, and Stewart}]{zhao2023rel2graph}
\bibinfo{author}{Z.~Zhao}, \bibinfo{author}{W.~Liu}, \bibinfo{author}{T.~French}, \bibinfo{author}{M.~Stewart},
\newblock \bibinfo{title}{Rel2graph: Automated mapping from relational databases to a unified property knowledge graph},
\newblock \bibinfo{journal}{arXiv preprint arXiv:2310.01080}  (\bibinfo{year}{2023}{\natexlab{c}}).
\bibitem[{Zhong et~al.(2024)Zhong, Zhong, Sun, Jin, Qin, and Zhang}]{zhong2024synthet2c}
\bibinfo{author}{Z.~Zhong}, \bibinfo{author}{L.~Zhong}, \bibinfo{author}{Z.~Sun}, \bibinfo{author}{Q.~Jin}, \bibinfo{author}{Z.~Qin}, \bibinfo{author}{X.~Zhang},
\newblock \bibinfo{title}{Synthet2c: Generating synthetic data for fine-tuning large language models on the text2cypher task},
\newblock \bibinfo{journal}{arXiv preprint arXiv:2406.10710}  (\bibinfo{year}{2024}).
\bibitem[{Ozsoy et~al.(2025)Ozsoy, Messallem, Besga, and Minneci}]{ozsoy2025text2cypher}
\bibinfo{author}{M.~G. Ozsoy}, \bibinfo{author}{L.~Messallem}, \bibinfo{author}{J.~Besga}, \bibinfo{author}{G.~Minneci},
\newblock \bibinfo{title}{Text2cypher: Bridging natural language and graph databases},
\newblock in: \bibinfo{booktitle}{Proceedings of the Workshop on Generative AI and Knowledge Graphs (GenAIK)}, \bibinfo{year}{2025}, pp. \bibinfo{pages}{100--108}.
\bibitem[{Guo et~al.(2023)Guo, Du, Liu, Zhou, He, and Han}]{guo2023gpt4graph}
\bibinfo{author}{J.~Guo}, \bibinfo{author}{L.~Du}, \bibinfo{author}{H.~Liu}, \bibinfo{author}{M.~Zhou}, \bibinfo{author}{X.~He}, \bibinfo{author}{S.~Han},
\newblock \bibinfo{title}{Gpt4graph: Can large language models understand graph structured data? an empirical evaluation and benchmarking},
\newblock \bibinfo{journal}{arXiv preprint arXiv:2305.15066}  (\bibinfo{year}{2023}).
\bibitem[{Xu et~al.(2024)Xu, Zhang, Jin, Zhu, Wu, and Weng}]{xu2024topochat}
\bibinfo{author}{H.~Xu}, \bibinfo{author}{B.~Zhang}, \bibinfo{author}{Z.~Jin}, \bibinfo{author}{T.~Zhu}, \bibinfo{author}{Q.~Wu}, \bibinfo{author}{H.~Weng},
\newblock \bibinfo{title}{Topochat: Enhancing topological materials retrieval with large language model and multi-source knowledge},
\newblock \bibinfo{journal}{arXiv preprint arXiv:2409.13732}  (\bibinfo{year}{2024}).
\bibitem[{Baraki(2024)}]{baraki2024leveraging}
\bibinfo{author}{W.~W. Baraki}, \bibinfo{title}{Leveraging large language models for accurate Cypher query generation: Natural language query to Cypher statements}, \bibinfo{type}{Master degree project}, University of Skövde, \bibinfo{year}{2024}. \bibinfo{note}{\url{https://www.diva-portal.org/smash/get/diva2:1881385/FULLTEXT01.pdf}}.
\bibitem[{Liu et~al.(2024)Liu, Wang, Ge, Wang, Xu, and Jia}]{liu2150text}
\bibinfo{author}{Y.~Liu}, \bibinfo{author}{X.~Wang}, \bibinfo{author}{J.~Ge}, \bibinfo{author}{H.~Wang}, \bibinfo{author}{D.~Xu}, \bibinfo{author}{Y.~Jia},
\newblock \bibinfo{title}{Text to graph query using filter condition attributes},
\newblock \bibinfo{journal}{Proceedings of the VLDB Endowment. ISSN} \bibinfo{volume}{2150} (\bibinfo{year}{2024}) \bibinfo{pages}{8097}.
\bibitem[{Liang et~al.(2024)Liang, Tan, Xie, Tao, Wang, Lan, and Qian}]{liang2024aligning}
\bibinfo{author}{Y.~Liang}, \bibinfo{author}{K.~Tan}, \bibinfo{author}{T.~Xie}, \bibinfo{author}{W.~Tao}, \bibinfo{author}{S.~Wang}, \bibinfo{author}{Y.~Lan}, \bibinfo{author}{W.~Qian},
\newblock \bibinfo{title}{Aligning large language models to a domain-specific graph database for nl2gql},
\newblock in: \bibinfo{booktitle}{Proceedings of the 33rd ACM International Conference on Information and Knowledge Management}, \bibinfo{year}{2024}, pp. \bibinfo{pages}{1367--1377}.
\bibitem[{Lei et~al.(2020)Lei, Wang, Ma, Gan, Lu, Kan, and Chua}]{lei2020re}
\bibinfo{author}{W.~Lei}, \bibinfo{author}{W.~Wang}, \bibinfo{author}{Z.~Ma}, \bibinfo{author}{T.~Gan}, \bibinfo{author}{W.~Lu}, \bibinfo{author}{M.-Y. Kan}, \bibinfo{author}{T.-S. Chua},
\newblock \bibinfo{title}{Re-examining the role of schema linking in text-to-sql},
\newblock in: \bibinfo{booktitle}{Proceedings of the 2020 Conference on Empirical Methods in Natural Language Processing (EMNLP)}, \bibinfo{year}{2020}, pp. \bibinfo{pages}{6943--6954}.
\bibitem[{Cao et~al.(2024)Cao, Zheng, Fan, Zhang, Chen, and Bai}]{cao2024rsl}
\bibinfo{author}{Z.~Cao}, \bibinfo{author}{Y.~Zheng}, \bibinfo{author}{Z.~Fan}, \bibinfo{author}{X.~Zhang}, \bibinfo{author}{W.~Chen}, \bibinfo{author}{X.~Bai},
\newblock \bibinfo{title}{Rsl-sql: Robust schema linking in text-to-sql generation},
\newblock \bibinfo{journal}{arXiv preprint arXiv:2411.00073}  (\bibinfo{year}{2024}).
\bibitem[{Maamari et~al.(2024)Maamari, Abubaker, Jaroslawicz, and Mhedhbi}]{maamari2024death}
\bibinfo{author}{K.~Maamari}, \bibinfo{author}{F.~Abubaker}, \bibinfo{author}{D.~Jaroslawicz}, \bibinfo{author}{A.~Mhedhbi},
\newblock \bibinfo{title}{The death of schema linking? text-to-sql in the age of well-reasoned language models},
\newblock \bibinfo{journal}{arXiv preprint arXiv:2408.07702}  (\bibinfo{year}{2024}).
\bibitem[{Yu et~al.(2018)Yu, Li, Zhang, Zhang, and Radev}]{yu2018typesql}
\bibinfo{author}{T.~Yu}, \bibinfo{author}{Z.~Li}, \bibinfo{author}{Z.~Zhang}, \bibinfo{author}{R.~Zhang}, \bibinfo{author}{D.~Radev},
\newblock \bibinfo{title}{Typesql: Knowledge-based type-aware neural text-to-sql generation},
\newblock \bibinfo{journal}{arXiv preprint arXiv:1804.09769}  (\bibinfo{year}{2018}).
\bibitem[{Dong et~al.(2019)Dong, Sun, Liu, Lou, and Zhang}]{dong2019data}
\bibinfo{author}{Z.~Dong}, \bibinfo{author}{S.~Sun}, \bibinfo{author}{H.~Liu}, \bibinfo{author}{J.-G. Lou}, \bibinfo{author}{D.~Zhang},
\newblock \bibinfo{title}{Data-anonymous encoding for text-to-sql generation},
\newblock in: \bibinfo{booktitle}{Proceedings of the 2019 Conference on Empirical Methods in Natural Language Processing and the 9th International Joint Conference on Natural Language Processing (EMNLP-IJCNLP)}, \bibinfo{year}{2019}, pp. \bibinfo{pages}{5405--5414}.
\bibitem[{Bogin et~al.(2019)Bogin, Gardner, and Berant}]{bogin2019global}
\bibinfo{author}{B.~Bogin}, \bibinfo{author}{M.~Gardner}, \bibinfo{author}{J.~Berant},
\newblock \bibinfo{title}{Global reasoning over database structures for text-to-sql parsing},
\newblock \bibinfo{journal}{arXiv preprint arXiv:1908.11214}  (\bibinfo{year}{2019}).
\bibitem[{Wang et~al.(2019)Wang, Shin, Liu, Polozov, and Richardson}]{wang2019rat}
\bibinfo{author}{B.~Wang}, \bibinfo{author}{R.~Shin}, \bibinfo{author}{X.~Liu}, \bibinfo{author}{O.~Polozov}, \bibinfo{author}{M.~Richardson},
\newblock \bibinfo{title}{Rat-sql: Relation-aware schema encoding and linking for text-to-sql parsers},
\newblock \bibinfo{journal}{arXiv preprint arXiv:1911.04942}  (\bibinfo{year}{2019}).
\bibitem[{Pourreza and Rafiei(2023)}]{pourreza2023din}
\bibinfo{author}{M.~Pourreza}, \bibinfo{author}{D.~Rafiei},
\newblock \bibinfo{title}{Din-sql: Decomposed in-context learning of text-to-sql with self-correction},
\newblock \bibinfo{journal}{Advances in Neural Information Processing Systems} \bibinfo{volume}{36} (\bibinfo{year}{2023}) \bibinfo{pages}{36339--36348}.
\bibitem[{Talaei et~al.(2024)Talaei, Pourreza, Chang, Mirhoseini, and Saberi}]{talaei2024chess}
\bibinfo{author}{S.~Talaei}, \bibinfo{author}{M.~Pourreza}, \bibinfo{author}{Y.-C. Chang}, \bibinfo{author}{A.~Mirhoseini}, \bibinfo{author}{A.~Saberi},
\newblock \bibinfo{title}{Chess: Contextual harnessing for efficient sql synthesis},
\newblock \bibinfo{journal}{arXiv preprint arXiv:2405.16755}  (\bibinfo{year}{2024}).
\bibitem[{Glass et~al.(2025)Glass, Eyceoz, Subramanian, Rossiello, Vu, and Gliozzo}]{glass2025extractive}
\bibinfo{author}{M.~Glass}, \bibinfo{author}{M.~Eyceoz}, \bibinfo{author}{D.~Subramanian}, \bibinfo{author}{G.~Rossiello}, \bibinfo{author}{L.~Vu}, \bibinfo{author}{A.~Gliozzo},
\newblock \bibinfo{title}{Extractive schema linking for text-to-sql},
\newblock \bibinfo{journal}{arXiv preprint arXiv:2501.17174}  (\bibinfo{year}{2025}).
\bibitem[{Yuan et~al.(2025)Yuan, Chen, Hong, Zhang, Huang, and Huang}]{yuan2025knapsack}
\bibinfo{author}{Z.~Yuan}, \bibinfo{author}{H.~Chen}, \bibinfo{author}{Z.~Hong}, \bibinfo{author}{Q.~Zhang}, \bibinfo{author}{F.~Huang}, \bibinfo{author}{X.~Huang},
\newblock \bibinfo{title}{Knapsack optimization-based schema linking for llm-based text-to-sql generation},
\newblock \bibinfo{journal}{arXiv preprint arXiv:2502.12911}  (\bibinfo{year}{2025}).
\bibitem[{Hart et~al.(1968)Hart, Nilsson, and Raphael}]{hart1968formal}
\bibinfo{author}{P.~E. Hart}, \bibinfo{author}{N.~J. Nilsson}, \bibinfo{author}{B.~Raphael},
\newblock \bibinfo{title}{A formal basis for the heuristic determination of minimum cost paths},
\newblock \bibinfo{journal}{IEEE transactions on Systems Science and Cybernetics} \bibinfo{volume}{4} (\bibinfo{year}{1968}) \bibinfo{pages}{100--107}.
\bibitem[{Liang et~al.(2024)Liang, Xie, Peng, Huang, Lan, and Qian}]{liang2024nat}
\bibinfo{author}{Y.~Liang}, \bibinfo{author}{T.~Xie}, \bibinfo{author}{G.~Peng}, \bibinfo{author}{Z.~Huang}, \bibinfo{author}{Y.~Lan}, \bibinfo{author}{W.~Qian},
\newblock \bibinfo{title}{Nat-nl2gql: A novel multi-agent framework for translating natural language to graph query language},
\newblock \bibinfo{journal}{arXiv preprint arXiv:2412.10434}  (\bibinfo{year}{2024}).
\bibitem[{HuggingFace(2024)}]{hfEvaluate}
\bibinfo{author}{HuggingFace}, \bibinfo{title}{Huggingface evaluate}, \bibinfo{year}{2024}. \bibinfo{note}{\url{https://huggingface.co/evaluate-metric}}.

\end{thebibliography}

\appendix


\end{document}